\documentclass[a4paper, oneside, twocolumn, notitlepage, 10pt]{extarticle_ecoc}
\usepackage{ecoc}

\usepackage{xpatch}

\makeatletter
\patchcmd\blx@bblinput{\blx@blxinit}
                      {\blx@blxinit
                      }{}{\fail}
\makeatother
			     
\begin{document}
\selectlanguage{english}

\title{Experimental Investigation into Split Nonlinearity Compensation in Single and Multi-channel WDM Systems}

\author{
    Ronit Sohanpal\textsuperscript{$*$}, 
    Eric Sillekens,
    Jiaqian Yang,
    R\^{o}mulo Aparecido, \\
    Zhixin Liu, 
    Robert Killey and
    Polina Bayvel
}

\maketitle

\begin{strip}
    \begin{author_descr}

        Optical Networks Group, UCL (University College London), London, UK. \textsuperscript{$*$}\textcolor{blue}{\uline{ronit.sohanpal@ucl.ac.uk}} 

    \end{author_descr}
\end{strip}

\renewcommand\footnotemark{}
\renewcommand\footnoterule{}

\begin{strip}
    \begin{ecoc_abstract}
        We experimentally investigated the performance of split nonlinearity compensation schemes for single and multi-channel WDM systems. We show that split NLC SNR gains of more than 0.4~dB at 5540~km can be achieved compared to transmitter- or receiver-side DBP alone when signal-ASE beating limits transmission performance.
        \textcopyright2024 The Author(s)
    \end{ecoc_abstract}
\end{strip}

\section{Introduction}

Digital nonlinearity compensation techniques have been widely investigated to increase the achievable signal-to-noise ratio (SNR) of long-reach coherent transmission systems. The digital backpropagation (DBP) scheme, which digitally propagates the signal in an inverted virtual channel to reverse accumulated chromatic dispersion and deterministic signal-signal nonlinear interactions, has been shown to be effective in mitigation of transmission impairments \cite{Ip2010,Ellis2017,Maher2015,Millar2010,Liga2014,Xu2017}. DBP can be conducted at either the transmitter side (known as Tx-DBP) or the receiver side (Rx-DBP). Previous work showed that Tx-DBP can reduce the accumulation of detrimental signal-amplified spontaneous emission (ASE) beating to achieve a single span improvement in transmission distance compared to Rx-DBP \cite{Lavery2017}.

It has previously been shown, theoretically, that splitting the nonlinearity compensation between the Tx and Rx, known as split nonlinearity compensation (split NLC), can achieve more than 1.5~dB gain in SNR \cite{Lavery2016}. This improvement arises because split NLC minimises the contribution of signal-ASE beating to $N/2$ spans, rather than $N$ spans for Rx-DBP ($N-1$ spans for Tx-DBP).
Further, investigations have shown that transceiver noise plays a key role in determining the effectiveness of split NLC at relatively short distances  \cite{Semrau2018}. Tx and Rx (TRX) noise  contributions adversely affect Tx-DBP and Rx-DBP gains respectively, with split NLC gain dependant on the total TRX noise as well as Tx/Rx noise distribution. TRX noise here is defined as that which determines back-to-back performance, e.g. quantisation noise from digital-to-analogue/analogue-to-digital converters, noise from linear electrical amplifiers and optical pre-amplifiers/boosters. Limited peak-to-average-power ratio (PAPR) of the Tx can be considered as additional transmitter noise for Tx-DBP due to the high PAPR of dispersion pre-compensation.

However, all prior studies on split NLC have thus far been theoretical, with all experimental DBP demonstrations shown for either Tx- or Rx-DBP only \cite{Temprana2015,Galdino2017}. Experimental implementation of split NLC impairments such as transceiver roll-off, gain tilt and inter-channel delays can cause a significant reduction of split NLC gain compared to theoretical predictions. While PMD and equalisation-enhanced phase noise have been included in prior theoretical studies, the overall conclusions on split NLC performance have yet to be shown experimentally \cite{Lavery2017}.

In this work we investigate for the first time the nonlinearity mitigation performance of split NLC via single and multi-channel coherent transmission experiments using a recirculating fibre loop. We studied the impact of both split ratio and transmission distance to determine the limits of signal-ASE and signal-TRX noise beating on the optimal SNR.

\section{Experimental setup}

\begin{figure*}[ht!]
  \centering
  \includegraphics[width=\textwidth]{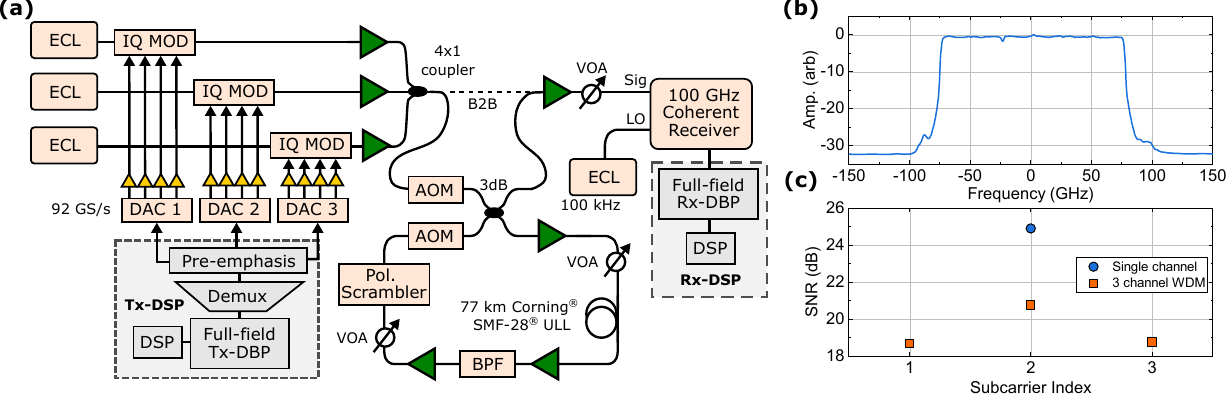}
\caption{(a) Experimental setup of the re-circulating loop based transmission. A 3x49.5~GBd 50-GHz-spaced DP-64QAM superchannel is back-propagated at either the transmitter or receiver and launched into multiple spans of 77~km ULL fibre. (b) Transmitted superchannel spectrum (0.02~nm resolution). (c) Back-to-back SNR for 1x49.5~GBd and 3x49.5~GBd signals.}
\label{fig_1}
\end{figure*}

The experimental setup is shown in Fig. \ref{fig_1}(a). Three C-band 100~kHz linewidth external cavity lasers (ECLs) located on a 50-GHz-spaced grid (centred at 1553~nm) were used to seed three dual-polarisation (DP) IQ modulators, each with 40~GHz 3-dB bandwidth. Each modulator was driven by a separate 92~GS/s arbitrary waveform generator (AWG) with 32~GHz analogue bandwidth and 8-bit vertical resolution (5~bit ENOB). Three pseudorandom data streams were used to generate three independent channels of pilot-based 49.5~GBd DP-64QAM signals shaped with 1\%-roll-off root-raised cosine filter. For full-field Tx-DBP, each data channel was multiplexed and digitally back-propagated through multiple spans of 76.96~km of Corning\textsuperscript{{\textregistered}} SMF-28\textsuperscript{{\textregistered}} ultra-low loss (ULL) fibre with a span loss of 12.2~dB using the symmetric split-step Fourier method with 1000 steps (77~m step size). The backpropagated signal was then demultiplexed into its subchannels and loaded onto the corresponding AWGs. For both single and multi-channel transmission, the edge channels (centre for single-channel) used the entire AWG bandwidth of 90~GHz to include any four-wave mixing sidebands produced by Tx-DBP on either side of the backpropagated signal \cite{Xu2017}. To minimise the transceiver roll-off penalty on Tx-DBP performance, digital pre-emphasis was applied to each subchannel before loading \cite{geiger2023}. The transmitted subchannels were amplified and combined into a 150~GBd superchannel.

\begin{figure*}[hb!]
  \centering
\includegraphics[width=\textwidth]{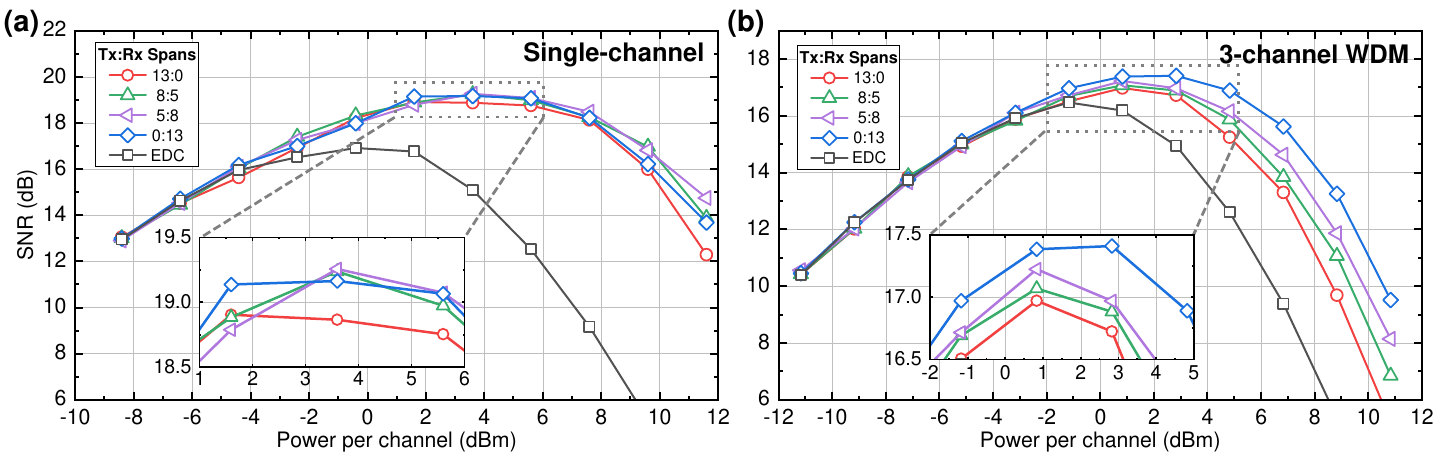}
\caption{SNR vs launch power per channel for (a) 1x49.5~GBd
and (b) 3x49.5~GBd signals over 1000.6~km (13x77.6~km spans) with either Tx-DBP (13:0, red circles), Rx-DBP (0:13, blue diamonds) or split.}
\label{fig_2}
\end{figure*}

The recirculating loop consisted of two acousto-optic modulators (AOMs) as switches to control loading and recirculation, a span of the aforementioned ULL fibre, a bandpass filter, polarisation scrambler and three 5-dB noise figure EDFAs for launch power control and optical power balancing in the loop.
For the back-to-back measurements, the signal was passed to a 200-GHz-bandwidth coherent receiver connected to a 256~GS/s 110~GHz Keysight UXR multi-channel oscilloscope with a 10-bit ADC (about 5-bit ENOB at 70~GHz). A 100-kHz local oscillator (LO) was aligned to the centre channel. All three subchannels were received simultaneously by a single receiver and each subchannel was digitally downconverted to baseband before passing through the DSP chain. For all scenarios considered, pilot-based DSP was used with a pilot sequence length of $2^{10}$ and a pilot rate of $1/32$ \cite{Wakayama2021}. For full-field Rx-DBP, the entire received superchannel was backpropagated using the same parameters as described for Tx-DBP. For both Tx- and Rx-DBP no electronic dispersion compensation (EDC) was used. To ensure all the Tx-DBP subchannels were temporally aligned, the signal was received back-to-back and the autocorrelation of each subchannel was calculated and used to delay the AWGs, aligning the subchannels to within 1 received sample (3.91~ps).

\begin{figure}[hb!]
  \centering \includegraphics[width=0.45\textwidth]{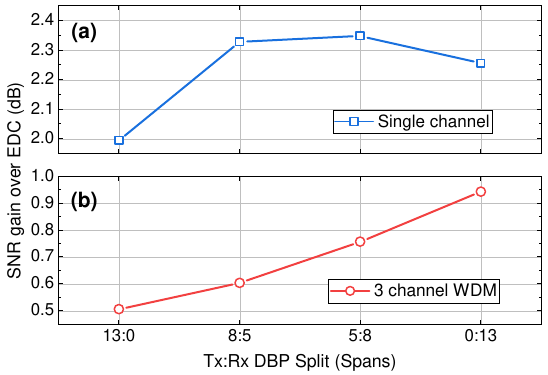}
\caption{Optimum SNR gain over EDC for (a) 1x49.5GBd and (b) 3x49.5~GBd signals versus split NLC ratio for 1000~km}
\label{fig_3}
\end{figure}

\section{Results}

The back-to-back SNR was measured for each channel and the results are shown in Fig. \ref{fig_1}(b) for both single-channel and three-channel transmission. For the single-channel, only the centre channel of the WDM super-channel was transmitted. In the three-channel transmission, the 2-dB lower SNR for the two edge channels can be attributed to the limited ENOB of the receiver. In all subsequent launch power measurements, only the centre channel SNR is shown.

The SNR versus launch power per channel was measured over 13 span (1000.6~km) transmission for single-channel and three-channel transmission and the results are shown in Fig. \ref{fig_2}(a) and Fig. \ref{fig_2}(b), respectively for different split NLC spans, ranging from full Tx-DBP (13:0) to full Rx-DBP (0:13). Fig. \ref{fig_3}(a) and Fig. \ref{fig_3}(b) show the peak SNR gain over EDC versus split NLC spans for single-channel and three-channel scenarios respectively.

For the single-channel transmission, the performance of all schemes is comparable, achieving approximately 2~dB SNR gain over EDC at optimal launch power. The largest gain of 2.35~dB occurs at a 5:8 split ratio, outperforming Tx-DBP by 0.35~dB and Rx-DBP by 0.09~dB. Short distance NLC performance is dominated by signal-TRX noise beating, thus split NLC gains are expected to be small relative to Tx- or Rx-DBP.

 For the three-channel case, the SNR gain is reduced to less than 1~dB for all scenarios, due to the aforementioned TRX noise beating. The highest SNR gain is achieved for Rx-DBP only. This indicates significantly greater Tx noise than Rx noise, which penalises Tx-DBP performance. This is due to the additional Tx noise contributions from 3 independent transmitters. In addition, the frequency-uncorrelated laser sources will greatly reduce Tx-DBP performance, particularly when dither-based wavelength tracking is used \cite{Sohanpal2023}.

\begin{figure*}[ht!]
  \centering
  \includegraphics[width=\textwidth]{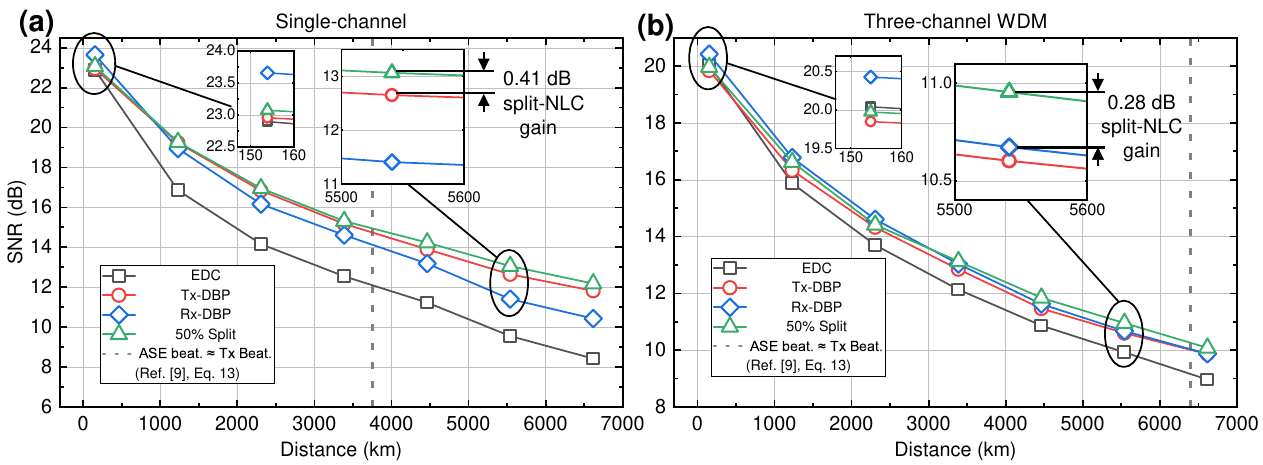}
\caption{Optimum SNR vs distance for (a) single-channel and (b) three channel WDM for EDC, Tx-DBP, Rx-DBP and 50\% split NLC.}
\label{fig_4}
\end{figure*}

Fig. \ref{fig_4} shows the optimum SNR as a function of the transmission distance for both single-channel (Fig. \ref{fig_4}(a)) and three-channel (Fig. \ref{fig_4}(b)) systems for EDC, full-field Tx-DBP, Rx-DBP and 50\% split NLC (i.e. 50\% spans at the transmitter and the receiver). The SNR corresponds to the peak SNR for optimum launch power. The dashed line shows the theoretical estimated distance at which TRX and ASE beating contributions are equal, derived from Eq. (13) in ref. [9] for a 50\% split. In both transmission scenarios, at short distances (500 to 1000~km) Rx-DBP gives the greatest SNR gain, suggesting dominant transceiver noise contributions to the TRX noise are penalising Tx-DBP and split NLC. As the transmission distance is increased, the signal-ASE beating increases, leading to a slow growth in the split NLC SNR gains relative to Tx- and Rx-DBP.

For the single-channel case, split NLC outperforms both Tx-DBP and Rx-DBP above 1200~km (16 spans), obtaining 0.41~dB gain over Tx-DBP and 1.75~dB gain over Rx-DBP at 5540~km. This suggests that signal-ASE beating dominates the optimal SNR performance instead of signal-transceiver beating. This behaviour is in strong agreement with the previously published theoretical analysis of split NLC in the presence of transceiver noise \cite{Semrau2018}. 

The same behaviour is observed for three channels except split NLC outperforms Tx-/Rx-DBP above 3400~km, achieving 0.28~dB gain over Rx-DBP and 0.38~dB gain over Tx-DBP. The significant Tx noise in the multichannel case increases the distance at which the signal-ASE dominates the SNR performance, and thus the distance at which split NLC gains occur. Note that the overall DBP gains are reduced compared to the single-channel scenario - this may be due to the larger overall TRX noise of multichannel transmission and single-receiver detection. Part of this excess TRX noise may be mitigated using a frequency comb, for which the mutual frequency coherence can minimise frequency-error-induced Tx-DBP penalties \cite{Temprana2015,Sohanpal2023}.
\section{Conclusions}

In this paper, we demonstrate the first experimental investigation into split NLC performance for both single and multi-channel coherent transmission systems. We show that transceiver noise is the dominant impairment to split NLC SNR gains at distances less than 3500~km. We find that split NLC outperforms both Tx- and Rx- by 0.41~dB in the long-distance signal-ASE beating regime, verifying theoretical predictions of split NLC behaviour.

\section{Acknowledgements}

The authors acknowledge EPSRC grants EP/R035342/1 TRANSNET (Transforming networks - building an intelligent optical infrastructure), EP/V007734/1 EPSRC Strategic Equipment Grant, EP/W015714/1 EWOC (Extremely Wideband Optical Fibre Communication Systems) and EP/V051377/1 ORBITS (Overcoming Resolution and Bandwidth Limit in Radio-Frequency Signal Digitisation). We also thank Sergejs Makovejs from Corning for providing the fibre used in this work.

\printbibliography[]

\vspace{-4mm}

\end{document}